\documentclass[12pt]{article}
\usepackage{amstex}
\begin{document}
\begin{center}
\textbf{ ENERGY  PRODUCTION  IN  THE  FORMATION  OF  A  FINITE  THICKNESS  COSMIC  STRING}\\
\bigskip
\bigskip
\bigskip
\bigskip
\bigskip
\bigskip
I. Brevik\footnote{E-mail address:  iver.h.brevik\@ mtf.ntnu.no}\\
{\it Division of Applied Mechanics,
Norwegian University of Science and Technology,\\
N-7491 Trondheim, Norway}\\
\bigskip
\bigskip
\bigskip
A. G. Fr{\o}seth\footnote{E-mail address:  anders.froseth\@ phys.ntnu.no}\\
{\it Department of Physics,
Norwegian University of Science and Technology,\\
N-7491 Trondheim, Norway}
\bigskip
\bigskip
September 1999
\end{center}
\bigskip
\begin{abstract}

The classical electromagnetic modes outside a long, straight, superconducting cosmic string are calculated, assuming the string to be surrounded by a superconducting cylindric surface of radius $R$. Thereafter, by use of a Bogoliubov-type argument, the electromagnetic energy $W$ produced per unit length in the lowest two modes is calculated when the string is formed "suddenly". The essential new element in the present analysis as compared with prior work of Parker [Phys. Rev. Lett. {\bf 59}, 1369 (1987)] and Brevik and Toverud [Phys. Rev. D {\bf 51}, 691 (1995)], is that the radius {\it a} of the string is assumed finite, thus necessitating Neumann functions to be included in the fundamental modes. We find that the theory is changed significantly: {\it W} is now strongly concentrated in the lowest mode $(m,s)=(0,1)$, whereas the proportionality $W \propto (G\mu /t)^2$ that is characteristic for zero-width strings is found in the next mode (1,1). Here $G$ is the gravitational constant, $\mu$ the string mass per unit length, and $t$ the GUT time.
\end{abstract}

\newpage
\section{Introduction}

Cosmic strings are remnants of an ultrahigh-temperature phase transition in the early universe, allowed in certain types of field theories \cite{vilenkin94}. They are explained in general terms of spontaneous symmetry breaking and have analogies to vortex lines in superfluid helium and in superconductors. If they exist, they may have seeded galaxy formation and may explain the observed fluctuations in the cosmic background radiation.

An interesting aspect of cosmic string theory is the estimate of the amount of radiation produced when the string is formed "suddenly", at some instant $t=t_0$, where $t_0 \simeq 10^{-35}$s  is the characteristic time for the grand unified field theory (GUT). One can express the number of massless particles created by the formation of the string in terms of Bogoliubov coefficients relating the initial metric to the (static) metric after the string is formed. This kind of approach was pioneered by Parker \cite{parker87}, for the case of a massless scalar field. As the radius $a$ of a string is believed to be very small ($\simeq 10^{-35}$ cm for symmetry breaking at the GUT scale) the string was in the formalism put equal to zero, thus making it possible to avoid the terms in the solution of the field equations that diverge at the centre ($r=0$). If factors of unity are disregarded, the analysis of Parker shows that the energy produced per unit length in the lowest non vanishing radiation mode is of order $W \sim (G\mu /t)^2$, where $\mu \simeq 10^{22}$ g/cm  is the string mass per unit length (GUT scale).

An analogous calculation to that of Parker was made in Ref. \cite{brevik95} in the electromagnetic case, still under the assumption that the string radius $a$ is equal to zero. This calculation is pertinent to the particular subclass of {\it superconducting} strings \cite{witten85}, believed to be existent in the early Universe under certain conditions. We shall not enter into a discussion of the superconducting string's physical properties here. Rather, we take a pragmatic attitude and adopt a physical model in which the string can be taken to obey perfect conducting boundary conditions at $r=a$. This kind of model is motivated by the properties known from electromagnetic theory for a superconductor \cite{landau84}: The magnetic field must be everywhere tangential to the surface, and the electric field must be everywhere normal to the surface. Thus we have precisely the same boundary conditions as if the surface were perfectly conducting.

In \cite{brevik95} we found essentially the same result as Parker did in the scalar case: For the lowest non vanishing mode, the energy $W$ was found to be of order $ (G\mu/t)^2$.

The purpose of the present paper is to calculate, again with use of the quantum mechanical sudden approximation, the amount of electromagnetic energy produced if the string has a {\it finite} radius $a$. The limit $a\rightarrow 0$ is more delicate than one might anticipate at first: The Neumann functions present in the formalism for finite $a$ blow up in this limit, and one cannot in advance know what the physical implications of this are, whether the emitted energy $W$ diverges, stays finite, or goes to zero. A concrete calculation is needed. 

In the next section we give the exterior Gott-Hiscock metric, which is a central element in the subsequent analysis. Thereafter, in Sec. 3 we calculate the three classical modes in the cavity, and the energy associated with the TM and TE modes. In the quantum mechanical theory in Secs. 4 and 5 we consider the TM mode only, and calculate by means of the Bogoliubov transformation the electromagnetic energy $W_{ms}$ produced in the mode $(m,s)$, for a GUT string. The energy production in the lowest mode $(0,1)$ is found to be large though finite: $W_{01} \simeq 6\times 10^{29}$ erg/cm. It is not until the next mode (1,1), where the energy is much smaller, that $W$ is found to obey the proportionality mentioned above: $W_{11}\propto (G\mu /t)^2$. The inclusion of the finite string width thus changes the theory in a significant way.

We do not consider the limit $a\rightarrow 0$ explicitly in this paper. We calculate however the emitted energy for a relative string thickness in the
interval $10^{-5} < a/R < 10^{-2}$. The result is that over this range the energy in the (0,1) mode is still dominant.

We adopt in the following Heaviside-Lorentz units, and put $\hbar = c= 1$.

\newpage
\section{The string metric}

In this section we collect some information about the metric around an infinitely long, straight string of finite radius $a$. Let the string's centerline coincide with the $z$-axis. We employ cylindrical coordinates, numerating them as $x^{\mu}=(t,r,\theta,z)$. 

We first note that within the conventional approach, when neglecting the string's width and using the linearized Einstein's equations, one has for the exterior metric
\begin{equation}
ds_{ext}^2= -dt^2+dr^2+(1-8G\mu)r^2d\theta^2+dz^2,
\end{equation}
\label{1}
where $G$ is the gravitational constant and $\mu$, as mentioned, is the string mass per unit length. The solution (1) was found by Vilenkin, within perturbation theory \cite{vilenkin81}. The deviation from the Minkowski metric is small, $G\mu \simeq 10^{-6}$.

Then let us take into account the finite radius $a$. We assume the mass density $\rho(r)$ to be uniformly distributed across the string,
\begin{equation}
\rho(r)= \left\{ \begin{array} {ll} 
                \rho_0 ~~~~~{\rm for}~~~ r \leq a,\\
                 0~~~~~~   {\rm for}~~~  r>a.
\end{array}
 \right.
\end{equation}
\label{2}
Then, with the abbreviation
\begin{equation}
\kappa= (8\pi G \rho_0)^{1/2},
\end{equation}
\label{3}
we can write the interior metric as
\begin{equation}
ds^2_{int}=-dt^2+dr^2+\left( \frac{\sin \kappa r}{\kappa}\right)^2 d\theta^2+dz^2,
\end{equation}
\label{4}
where $ -\infty < t < \infty, ~0<r<\theta_M/\kappa $  where $\theta_M$ is a constant, $0 \leq \theta < 2\pi$, and $-\infty < z < \infty$. 

To find the metric on the outside, we have to take into account the conditions that the metric itself, as well as its first derivative with respect to $r$, have to be continuous across the surface of the string (these are the Israel matching conditions \cite{israel66}).  We get
\begin{equation}
ds^2_{ext}= -dt^2+dr^2+ (1-4G\mu)^2 r^2 d\theta^2+ dz^2.
\end{equation}
\label{5}
This is the exact metric, valid for $r \geq a$, and will be used in the following. The constant $\theta_M$ introduced above is related to the mass per unit length by the relation
\begin{equation}
\mu= \frac{1}{4G}(1-\cos \theta_M).
\end{equation}
\label{6}
Strictly speaking, the metric (5) applies only for $G\mu < \frac{1}{4}$. But, as noted, this condition is amply satisfied for cosmic strings.

In the case of a weak field, $ \kappa a \ll 1$ which implies  $G\mu \ll 1$, the exact solution (5) is seen to reduce to Vilenkin's singular line metric (1). Then Eq. (6) yields $8G\mu=\theta_M^2$ which, together with $\mu=\pi a^2 \rho_0$, yields $\theta_M=\kappa a$. The upper limit $\theta_M/\kappa$ for $r$ in the interior metric (4) thus becomes equal to $a$.

The exact solutions  (4) and  (5) were found by Gott \cite{gott85}, and by Hiscock \cite{hiscock85}. Cf. also the papers of  Helliwell and Konkowski \cite{helliwell86}, Jensen \cite{jensen97},  and the book of Peebles \cite{peebles93}.

\section{The classical electromagnetic modes}

The superconducting string of radius $a$, as mentioned, is assumed to be formed suddenly at the instant $t=t_0$, of the order of the GUT time. We assume that the electromagnetic field, at all times, is constrained to lie within a large cylinder of length $L$ and radius $R$, whose centerline is coinciding with that of the string (the  $z$-axis). Here $R$ and $L$ are taken to be of the same order of magnitude as the mean spacing between strings at the time of formation. This length can be no larger than the horizon size, which in a radiation dominated universe is $R_p=2t$. We will use $R=t$. For the electromagnetic fields we assume perfect conductor boundary conditions, on the surface $r=R$ as well as on $r=a$.

Prior to the formation of the string we have the Minkowski metric
\begin{equation}
ds^2=-dt^2+dr^2+r^2d\theta^2+dz^2,
\end{equation}
\label{7}
where $t<t_0, ~ 0 \leq r \leq R, ~ 0\leq \theta<2\pi, ~-\frac{1}{2}L<z<\frac{1}{2}L$. After the formation, the metric is the Gott-Hiscock metric (5) in the exterior region. It is convenient to introduce the symbol
\begin{equation}
\beta=(1-4G\mu)^{-1}.
\end{equation}
\label{8}
The field modes calculated for $t<t_0$ are related to the modes for $t>t_0$ by a Bogoliubov transformation. The change in the metric has the effect that the vacuum states associated with the two sets of modes differ. As a consequence there will be a net particle production in the $t>t_0$ modes, relative to the $t<t_0$ vacuum. This is precisely the effect that is calculable in terms of the Bogoliubov coefficients of the transformation \cite{birrell82}. Before embarking on this dynamical calculation, we consider first the stationary classical modes.

\subsection{Modes for $t>t_0$}

Taking the time factor to be $\exp({-i\omega t})$ in a complex classical representation for the fields, we get Maxwell's equations in the annular region between $r=a$ and $r=R$:
\begin{equation}
{\bf \nabla \cdot E}=0, ~~~~{\bf \nabla \cdot H}=0,
\end{equation}
\label{9}
\begin{equation}
{\bf \nabla \times E}=i\omega {\bf H}, ~~~~{\bf \nabla \times H}=-i\omega {\bf E}.
\end{equation}
\label{10}
Expressed on component form, these equations are the same as Eqs. (5) - (8) in \cite{brevik95}, and will not be repeated here. From these equations we can derive a second order equation for $E_z$. We write this field component in the form $E_z(r,\theta,z)= E_z(r)\exp {i(m\theta+kz)}$, where $m$ is an  integer, and $k$ is the axial wave number. We assume that $k$ is continuous for $-\infty < k < \infty $. The equation for $E_z$ becomes
\begin{equation}
\left[ \frac{d^2}{dr^2}+\frac{1}{r}\frac{d}{dr}+\left( q^2-\frac{\beta^2 m^2}{r^2}\right) \right] E_z(r)=0,
\end{equation}
\label{11}
so that the solutions are linear combinations of ordinary Bessel functions $J_{|\beta m|}(qr)$ and Neumann functions  $N_{|\beta m |}(qr)$, with $q^2=\omega^2-k^2$. 

\bigskip

{\it TM mode:}

\bigskip

According to the boundary conditions, $E_z$ vanishes for $r=a, R$. From this we derive the condition
\begin{equation}
J_{|\nu|}(qR)N_{|\nu|}(qa)-N_{|\nu|}(qR)J_{|\nu|}(qa)=0,
\end{equation}
\label{12}
where we have defined $\nu$ as
\begin{equation}
\nu=\beta m.
\end{equation}
\label{13}
For a fixed value of $m$, Eq. (12) gives an infinite sequence of roots $q_{\nu s}$, with $s=1,2,3,...$ As $k$ is real, we can now write 
\begin{equation}
\omega_{\nu s k}=(q_{\nu s}^2+k^2)^{1/2}.
\end{equation}
\label{14}
We let $ A_{\nu s k}$ be an arbitrary mode-dependent constant, define $\cal{C}_{|\nu|}$ as
\begin{equation}
{\cal{C}}_{|\nu|}(q_{\nu s}r)=J_{|\nu|}(q_{\nu s}r)-\frac{J_{|\nu|}(q_{\nu s}a)}{N_{|\nu|}(q_{\nu s}a)}
N_{|\nu|}(q_{\nu s}r),
\end{equation}
\label{15}
and use the real representation for the azimuthal $\theta$ direction. The field components can then be written as
\begin{eqnarray}
E_r = A_{\nu s k}\frac{ik}{q_{\nu s}}\mathcal{C}_{|\nu|}'(q_{\nu s}r)\binom{\text{cos}}{\text{sin}}m\theta e^{i\Phi}, \\
E_{\theta} = \mp A_{\nu s k}\frac{ik\nu}{q_{\nu s}^2r}\mathcal{C}_{|\nu|}(q_{\nu s}r)\binom{\text{sin}}{\text{cos}}m\theta e^{i\Phi}, \\
E_z = A_{\nu s k}\mathcal{C}_{|\nu|}(q_{\nu s}r)\binom{\text{cos}}{\text{sin}}m\theta e^{i\Phi}, \\
H_r = \pm A_{\nu s k}\frac{i\omega_{\nu sk}\nu}{q_{\nu s}^2r}\mathcal{C}_{|\nu|}(q_{\nu s}r)\binom{\text{sin}}{\text{cos}}m\theta e^{i\Phi}, \\
H_{\theta} = A_{\nu s k}\frac{i\omega_{\nu sk}}{q_{\nu s}}\mathcal{C}_{|\nu|}'(q_{\nu s}r)\binom{\text{cos}}{\text{sin}}m\theta e^{i\Phi},
\end{eqnarray}
where $\Phi=kz-\omega_{\nu sk}t$. Prime means differentiation with respect to the whole argument. Note that the $m=0$ mode is independent of the azimuthal direction $\theta$.

Comparison with the analogous expressions given in \cite{brevik95} shows that the difference between the two cases lies in the radial functions. Explicitly, $J_{|\nu|} \rightarrow \cal{C}_{|\nu |}$, which in turn implies the presence of the Neumann function. The physical reason for this is, of course, the finiteness of the string's width.

\bigskip

{\it TE mode:}

\bigskip

This mode is constructed by starting from the governing equation for the component $H_z$. The boundary conditions now lead to the condition
\begin{equation}
J_{|\nu|}'(q'R)N_{|\nu|}'(q'a)-N_{|\nu|}'(q'R)J_{|\nu|}'(q'a)=0,
\end{equation}
\label{21}
which has roots $q_{\nu s}',¨s=1,2,3...$ The complete expression for the TE field components are 
\begin{eqnarray}
H_r = \tilde{A}_{\nu s k}\frac{ik}{q_{\nu s}'}\mathcal{C}_{|\nu|}'(q_{\nu s}'r)\binom{\text{cos}}{\text{sin}}m\theta e^{i\Phi'}, \\
H_{\theta} = \mp \tilde{A}_{\nu s k}\frac{ik\nu}{(q_{\nu s}')^2r}\mathcal{C}_{|\nu|}(q_{\nu s}'r)\binom{\text{sin}}{\text{cos}}m\theta e^{i\Phi'}, \\
H_z = \tilde{A}_{\nu s k}\mathcal{C}_{|\nu|}(q_{\nu s}'r)\binom{\text{cos}}{\text{sin}}m\theta e^{i\Phi'}, \\
E_r = \mp \tilde{A}_{\nu s k}\frac{i\omega_{\nu sk}'\nu}{(q_{\nu s}')^2r}\mathcal{C}_{|\nu|}(q_{\nu s}'r)\binom{\text{sin}}{\text{cos}}m\theta e^{i\Phi'}, \\
E_{\theta} = -\tilde{A}_{\nu s k}\frac{i\omega_{\nu sk}'}{q_{\nu s}'}\mathcal{C}_{|\nu|}'(q_{\nu s}'r)\binom{\text{cos}}{\text{sin}}m\theta e^{i\Phi'}, 
\end{eqnarray}
where the $\tilde{A}_{\nu sk}$ are a new set of constants, $\omega_{\nu sk}'=[(q_{\nu s}')^2+k^2]^{1/2}$, and $\Phi'=kz-\omega_{\nu sk}'t$.

\bigskip

{\it TEM mode:}

\bigskip

In this mode both components $E_z$ and $H_z$ are equal to zero. The field ${\bf E}(r,\theta)$ is the solution of a two-dimensional electrostatic problem, just as in the Minkowskian case \cite{choudhury89}. With ${\bf E}(r,\theta)=-{\bf \nabla}\phi$ inserted into ${\bf \nabla \cdot E}=0$ we get the generalized Poisson equation
\begin{equation}
\frac{1}{r}\frac{\partial}{\partial r}\left( r\frac{\partial \phi}{\partial r} \right) +\frac{\beta^2}{r^2}\frac{\partial^2 \phi}{\partial \theta^2}=0.
\end{equation}
\label{27}
The boundary conditions $E_r(a,\theta)=E_r(R,\theta)=0$ imply azimuthal independence of $\phi$, and we get the solution of Eq. (27) in the form $\phi=A-B\ln r$ with $A,B$ as constants. From Maxwell's equations we get the mode solutions
\begin{equation}
{\bf E}=\frac{B_k}{r}e^{i\Phi} \hat{r}, ~~~~{\bf H}(r)=\frac{B_k}{r}e^{i\Phi} \hat{\theta}.
\end{equation}
\label{28}
It turns out that the TEM mode is special and poses problems when developing a manageable quantum formalism for the field modes. This mode will therefore not be considered in the quantum theory below.

\subsection{Modes for $t<t_0$}

The initial conditions are those of an empty cylinder of radius $R$ in Minkowski space. We do not give the complete solution here. As in the case considered in \cite{brevik95}, the only components needed  for the calculation are $E_z$ and $H_z$ for the TM and TE modes, respectively. For $t<t_0$ these are
\begin{eqnarray}
E_z = B_{m s}J_{|m|}(p_{m s} r)\binom{\text{cos}}{\text{sin}}m\theta e^{i\Phi}, & \text{TM mode} \\
H_z = \tilde{B}_{m s k}J_{|m|}'(p_{m s}' r)\binom{\text{cos}}{\text{sin}}m\theta e^{i\Phi'},& \text{TE mode}
\end{eqnarray}
where $p_{ms},~s=1,2,3,...$ and $p_{ms}', ~s=1,2,3,...$ are roots of
\begin{equation}
J_{|m|}(pR)=0 ~~~~ \text{and}~~~~J'_{|m|}(p'R)=0.
\end{equation}
\label{31}
The TEM mode does not exist for $t<t_0$. This is just similar to the Minkowskian case.

\subsection{Energy considerations}

We note first the following point, without going into detailed considerations:
Using Maxwell's equations, we can find out whether there is an axial electric current $I_z$ flowing in the string, in the TE and TM modes. Similarly we can find out whether there is a charge $Q/L$ per unit length. Substituting the appropriate field components into the expressions for $I_z$ and $Q/L$ we find that it is only the mode $m=0$ that gives an answer different from zero. A net current, and an electric charge, are thus associated with azimuthal symmetry only. 

We next concentrate on the total electromagnetic energy per unit length, $W$, outside the string:
\begin{equation}
W=\frac{1}{4}\int _a^R (|E|^2+|H|^2)dA,
\end{equation}
\label{32}
where $dA=\beta^{-1}r dr d\theta$ is the cross-sectional area element of the guide. We get
\begin{equation}
W=
\begin{cases}
	\frac{\omega_{\nu sk}^2}{2q_{\nu s}^4}\int|\nabla_tE_z|^2dA, &                               \text{TM mode}\\
\\
	\frac{(\omega_{\nu sk}')^2}{2(q_{\nu s}')^4}\int|\nabla_tH_z|^2dA, &                                     \text{TE mode}
\end{cases}
\end{equation}
where ${\bf \nabla}_t$ is the transverse gradient which for the TM mode is defined by 
\begin{equation}
\int|\nabla_tE_z|^2dA \equiv \frac{1}{\beta}\int \left[\left|\frac{\partial E_z}{\partial r}\right|^2 + \frac{\beta^2}{r^2}\left|\frac{\partial E_z}{\partial \theta}\right|^2\right]rdrd\theta.
\end{equation}
Green's first identity together with the wave equation and boundary conditions for $E_z$ gives 
\begin{equation}
\int|\nabla_tE_z|^2dA=q_{\nu s}^2\int|E_z|^2dA,
\end{equation}
which, together with an analogous result for the TE mode, leads to the convenient result that the field energy is expressible entirely in terms of the $E_z$ and $H_z$ components:
\begin{equation} \label{energy}
W=
\begin{cases}
	\frac{\omega_{\nu sk}^2}{2q_{\nu s}^2}\int|E_z|^2dA, &                               \text{TM mode},\\
\\
	\frac{(\omega_{\nu sk}')^2}{2(q_{\nu s}')^2}\int|H_z|^2dA, &                                     \text{TE mode}.
\end{cases}
\end{equation}
Explicit calculations can now be carried out using the boundary conditions together with the orthogonality properties of the Bessel functions \cite{abramowitz72}. We get for the TM mode
\begin{equation}
W = \frac{\pi\omega_{\nu sk}^2}{2\beta q_{\nu s}^2}|A_{\nu sk}|^2[R^2\mathcal{C}_{|\nu|+1}^2(q_{\nu s}R) - a^2\mathcal{C}_{|\nu|+1}^2(q_{\nu s}a)], 
\end{equation}
and for the TE mode
\begin{equation}
\begin{split}
W &= \frac{\pi(\omega_{\nu sk}')^2}{2\beta (q_{\nu s}')^2}|A_{\nu sk}'|^2 \\
& \times[(R^2-\frac{\nu^2}{q_{\nu s}'})\mathcal{C}_{|\nu|}^2(q_{\nu s}'R) - (a^2-\frac{\nu^2}{q_{\nu s}'})\mathcal{C}_{|\nu|}^2(q_{\nu s}'a)]. \\
\end{split}
\end{equation}

\section{Quantum mechanical modes}

As in Ref. \cite{brevik95}, when developing the quantum formalism, we will concentrate on the TM mode only. We have seen that the field energy can be expressed solely in terms of the component $E_z$. We can therefore essentially make use of the formalism of scalar field theory. We expand $E_z$ as a field operator:
\begin{equation}
E_z = \int_{-\infty}^{\infty}\frac{dk}{2\pi}\sum_{\nu,s}[a_{\nu sk}u_{\nu sk}(x)+ a_{\nu sk}^{\dagger}u_{\nu sk}^*(x)],
\end{equation}
where $a_{\nu sk}$ and $a_{\nu sk}^{\dagger}$ satisfy the commutation relations
\begin{equation}
[a_{\nu sk},  a_{\nu' s' k'}^{\dagger}]=2\pi \delta(k-k')\delta_{\nu \nu'}\delta_{s s'},
\end{equation}
\label{40}
the other commutators vanishing. Here we use $\nu$ as a general index, meaning that
\begin{equation}
\nu =
\begin{cases}
m, &  \text{for $t<t_0$}\\
\beta m,& \text{for $t>t_0$}. 
\end{cases}
\end{equation}
The mode functions $u_{\nu sk}$ are normalized according to
\begin{equation}
\frac{\omega_{\nu sk}^2}{2q_{\nu s}^2}\int \langle |E_z|^2 \rangle dV=\langle a^\dagger a \rangle \omega_{\nu sk},
\end{equation}
\label{42}
where $dV$ is the volume element. This ensures that the quantum mechanical expression equivalent to (36) becomes equal to the expectation value of the number operator times the photon energy of the mode. Thus, the normalized mode functions for $t > t_0$ are
\begin{equation} \label{maft}
\begin{split}
u_{\beta msk}(x) =  &
\epsilon_m\sqrt{\frac{\beta}{\pi\omega_{\beta msk}}}\frac{q_{\beta ms}}{|R^2\mathcal{C}_{|\beta m|+1}^2(q_{\beta ms}R) -a^2\mathcal{C}_{|\beta m|+1}^2(q_{\beta ms}a)|^{1/2}} \\
& \times \mathcal{C}_{|\beta m|}(q_{\beta ms}r) \binom{\text{cos}}{\text{sin}}m\theta e^{i\Phi}, 
\end{split}
\end{equation}  
where 
\begin{equation}
\epsilon_m= \left\{ \begin{array}{ll}
1,~~~~m \neq 0 \\
\sqrt{2},~~m=0
\end{array}
\right.
\end{equation}
appears as a consequence of the real representation for the $\theta$ direction. The corresponding functions for $t<t_0$ are
\begin{equation} \label{mbef}
u_{msk}(x) =  \epsilon_m\sqrt{\frac{1}{\pi\omega_{msk}}}\frac{p_{ms}}{R|J_{|m|+1}(p_{m s}R)|} 
J_{|m|}(p_{ms}r)\binom{\text{cos}}{\text{sin}}m\theta e^{i\Phi}.
\end{equation}
We define the Klein-Gordon scalar product as in scalar field theory:
\begin{equation}
(\phi_1, \phi_2) = -\frac{i}{q^2}\int[\phi_1\partial_t\phi_2^*-(\partial_t\phi_1)\phi_2^*] dV,
\end{equation}
where $q^2=\sum_{\nu s} q_{\nu s}^2$. The mode functions are then orthonormal:
\begin{equation}
(u_{\nu sk}, u^*_{\nu' s' k'})=2\pi \delta(k-k')\delta_{\nu \nu'}\delta_{s s'},
\end{equation}
\label{47}
the other products vanishing.

For later calculations it will be important to specify the continuity conditions for the field at $t=t_0$. First, we require that the field itself is continuous:
\begin{equation}
E_z(x) |_{t_0^-}=E_z(x)|_{t_0^+}.
\end{equation}
\label{48}
Next, we can use the conservation of the Klein-Gordon product to find the continuity condition for $\partial_t E_z$:
\begin{equation}
\begin{split}
\frac{-i}{\beta q^2}\int[E_z|_{t_0^+}(\partial_tE_z^*)|_{t_0^+}-(\partial_tE_z)|_{t_0^+}E_z^*|_{t_0^+}]rdrd\theta dz= \\
\frac{-i}{p^2}\int[E_z|_{t_0^-}(\partial_tE_z^*)|_{t_0^-}-(\partial_tE_z)|_{t_0^-}E_z^*|_{t_0^-}]rdrd\theta dz.
\end{split}
\end{equation}
Substituting Eq. (48) into this equation we get
\begin{equation}
\partial_tE_z(x)|_{t_0^-}=\frac{p^2}{\beta q^2}\partial_t E_z(x)|_{t_0^+},
\end{equation}
\label{50}
where $p^2=\sum_{ms}p^2_{ms}$.

\section{Energy production calculated from the Bogoliubov coefficients}

\subsection{General formalism}

The two sets of mode functions give two different expansions for the field:  For $t<t_0$, $E_z$ is given by Eq. (39) with $\nu=m$; for $t>t_0$, $E_z$ is given by the same expression with $\nu=\beta m$. For each of these expressions there is a unique vacuum state:
\begin{equation}
a_{msk}|0\rangle_{msk}=0, ~~~\rm{for}~~ t<t_0,
\end{equation}
\label{51}
\begin{equation}
a_{\beta msk}|0\rangle_{\beta msk}=0, ~~~\rm{for}~~ t>t_0.
\end{equation}
\label{52}
As both sets of modes are complete, $u_{\beta msk}$ can be expanded in terms of $u_{msk}$:
\begin{equation}
\begin{split}
u_{\beta msk}(x)= & \int_{-\infty}^{\infty}\frac{dk'}{2\pi}\sum_{m',s'}[\gamma(\beta msk|m's'k')u_{m's'k'}(x) \\
&+ \delta(\beta msk|m's'k')u_{m's'k'}^*(x)],
\end{split}
\end{equation}
$\gamma$ and $\delta$ being the Bogoliubov coefficients \cite{birrell82}. Equivalently, for the operators we have the expansions
\begin{equation} \label{bogo1}
\begin{split} 
a_{\beta msk}= & \int_{-\infty}^{\infty}\frac{dk'}{2\pi}\sum_{m',s'}[\gamma(\beta msk|m's'k')a_{m's'k'} \\
&+ \delta^*(\beta msk|m's'k')a_{m's'k'}^{\dagger}].
\end{split}
\end{equation}
This implies that the average number of particles produced in the mode $m, s, k$ per unit interval in the $k$ space is
\begin{equation}
\frac{dN_{msk}}{dk} = \int_{-\infty}^{\infty}\frac{dk'}{2\pi}\sum_{m',s'}|\delta(\beta msk|m's'k')|^2.
\end{equation}
The coefficient $\delta(\beta msk|m' s' k')$ can be found using the continuity condition for the field together with the orthonormality of the scalar product:
\begin{equation} \label{bog}
\begin{split}
\delta(\beta msk|m's'k')= 
\frac{i}{\beta q_{\beta ms}^2}\int[u_{msk}(x)\partial_tu_{\beta m's'k'}(x) \\
-\beta\frac{q^2}{p^2}\partial_tu_{msk}(x)u_{\beta m's'k'}(x)]rdrd\theta dz.
\end{split}
\end{equation}
Substituting Eqs. (43) and (45) into Eq. (56) we have
\begin{equation} \label{energi}
\begin{split}
\delta(\beta msk|m's'k')& = \frac{2\pi}{\sqrt{\beta}} \frac{p_{msk}q_{\beta m's'k}}{q_{\beta m's'k'}^2}\delta_{mm'}\delta(k-k') \\
&\times\left[\sqrt{\frac{\omega_{\beta m's'k'}}{\omega_{msk}}}-\beta\frac{q^2}{p^2}
\sqrt{\frac{\omega_{msk}}{\omega_{\beta m's'k'}}}\right]I_{ss'}.
\end{split}
\end{equation}
where
\begin{equation}\label{integral}
I_{ss'} =
\frac{\int_a^RJ_{|m|}(p_{m s}r)\mathcal{C}_{|\beta m|}(q_{\beta ms'}r)rdr} 
{|RJ_{|m|+1}(p_{m s}R)||R^2\mathcal{C}_{|\beta m|+1}^2(q_{\beta ms'}R) - a^2\mathcal{C}_{|\beta m|+1}^2(q_{\beta ms'}a)|^{\frac{1}{2}}}.
\end{equation}
Our main task now is to calculate these expressions with satisfactory accuracy.

\subsection{Evaluation}

As $\beta$ is close to unity, we can put $\beta=1$ everywhere except in the difference between the square roots. We can also approximate the roots $q_{\beta ms}$ of the TM mode equation (12): As $a/R \sim 10^{-5}$ and the Neumann function diverges for small arguments, we can neglect the second term in the equation and so obtain $J_{|\beta m|}(q_{\beta m}R) \simeq 0$, which means that $q_{\beta ms} \simeq p_{ms}$. Using this, together with the general properties of the Bessel function, we find that Eq. (58) can be approximated by 
\begin{equation}
I_{ss'} \simeq \frac{1}{2}\delta_{ss'}.
\end{equation}
\label{59}
More detailed considerations on the evaluation of $I_{ss'}$ are given in Appendix A. Equation (57) can now be written
\begin{equation}
\delta(\beta msk|m' s' k' )=\pi \left[ \sqrt{\frac{\omega_{\beta m' s' k'}}{\omega_{msk}}}-
\beta\sqrt{\frac{\omega_{msk}}{\omega_{\beta m' s' k'}}} \right] \delta_{ss'}\delta_{mm'}\delta(k-k').
\end{equation}
\label{60}
This leads to the following expression for the electromagnetic energy production in the mode $m,s,k$ per unit length, and per unit $k$ interval:
\begin{equation}
\frac{dW_{msk}}{dk}
=\frac{\omega_{msk}}{L}\frac{dN_{msk}}{dk}=\frac{1}{4}\frac{(\omega_{\beta msk}-\beta \omega_{msk})^2}{\omega_{\beta msk}}.
\end{equation}
\label{61}
As in \cite{brevik95} and \cite{parker87} we simplify the calculations by considering only the wave number interval centered around $k=0$.  Since  $\omega_{\beta msk}=(q^2_{\beta ms}+k^2)^{1/2}$  and  $\omega_{msk}=(p^2_{ms}+k^2)^{1/2}$,  it follows that   $\omega_{\beta ms0}=q_{\beta ms}$   and   $ \omega_{ms0}=p_{ms}$.  Therefore, the energy production can be found from the roots of the Bessel function:
\begin{equation}
\frac{dW_{msk}}{dk}|_{k=0}=\frac{1}{4}\frac{(q_{\beta ms}-\beta p_{ ms})^2}{q_{\beta ms}}  .
\end{equation}
\label{62}
In order to make a reasonably accurate estimate of the energy production one has to be careful: As $|1-\beta| \simeq 4\times 10^{-6}$ it follows that insertion of crude approximations for the roots may easily lead to substantial errors. The best way to proceed, is to check the roots on a computer. In Table I we give a short compilation of roots  $x_{\beta ms}=Rq_{\beta ms}$  and  $x_{ms}=Rp_{ms}$  calculated by use of Maple V.

It is clear that in the lowest mode, $(m,s)=(0,1)$, the difference between the non gravitational and the gravitational cases is most significant. Let us compare the energy produced in this mode with that of the next mode, $(m,s)=(1,1)$:
\begin{equation}
{\frac{dW_{msk}}{dk}\vline}_{k=0} \simeq 
\begin{cases}
\frac{1}{4R}\times 10^{-2} \simeq \frac{2}{R}\times 10^{-3}, & \text{for $(0,1)$} \\
\frac{1}{R}(1-\beta)^2 \simeq \frac{2}{R}\times 10^{-11}, & \text{for $(1,1)$} . 
\end{cases}
\end{equation}
The tendency to produce more energy in the lowest mode is very strong: The value for $(0,1)$ is $10^8$ times larger than the value for $(1,1)$.

To get the total energy, we multiply the above equation with the width $\Delta k$ of wave numbers contributing to the energy production. As the string actually forms during a finite time interval $\Delta t$, the production of photons with wave numbers much larger than $1/\Delta t$ will be suppressed. We take $\Delta t =R=t$. For the total energy per unit length we then obtain
\begin{equation}
W_{ms} \simeq \frac{1}{t}{\frac{dW_{msk}}{dk}\vline}_{k=0} \simeq 
\begin{cases}
\frac{2}{t^2}\times 10^{-3}, & \text{for $(0,1)$} \\
\frac{2}{t^2}\times 10^{-11}, & \text{for $(1,1)$}.  
\end{cases}
\end{equation}
As mentioned above, we take $t$ to be the GUT time, $t \simeq 10^{-35}$ s. For the $(0,1) $ mode we then have, in ordinary units,
\begin{equation}
W_{01} \simeq  \frac{2 \hbar}{c^5t^2}\times 10^{-3} \simeq 6 \times 10^{29} \frac{\text{erg}}{\text{cm}},
\end{equation}
and for the $(1,1)$ mode,
\begin{equation}
W_{11} \simeq \frac{2 \hbar}{c^5t^2}\times 10^{-11} \simeq 6 \times 10^{21} \frac{\text{erg}}{\text{cm}}.
\end{equation}
Characteristic for the finite-width string considered here is that the dominant mode is the one corresponding to azimuthal symmetry, $m=0$. For the zero-width string considered in \cite{brevik95}, we got instead the emitted energy in this mode to be  {\it zero}. Inclusion of the finite width thus changes the theory in an essential way. The lowest non vanishing mode found in \cite{brevik95} was $(m,s)=(1,1)$. From Eqs. (63) and (64) we see that, in the present case, the $(1,1)$ mode corresponds to $W_{11}$ being proportional to $(G\mu/t)^2$. This agrees qualitatively with the behavior found in \cite{brevik95} for the same mode.  Quantitatively, Eq. (66) is of the same order of magnitude as Eq. (57) in \cite{brevik95}.

\subsection{On the $a/R$- dependence}

In view of the sensitivity that we have seen in the calculation of the energy production, we may ask to what extent the results are dependent on the values chosen for the input parameters.

Recall that our input parameters are the string mass $\mu$ per unit length, the string radius $a$, the string length $L$, and the radius $R$ of the waveguide. In terms of these parameters the energy production per unit length can be written as (cf. Eq. (61)):
\begin{equation}
W_{ms} = \frac{1}{4LR}\frac{[x_{\beta ms}-(1-4G\mu)^{-1}x_{ms}                                              ]^2}{x_{\beta ms}}.
\end{equation}
Here it is to be noted that the roots $x_{\beta ms}$ depend on the ratio $a/R$. Table II gives the roots for various values of $a/R$ (they were calculated using Maple V). Note also that the roots $x_{ms}$ are independent of $a$ and $R$, and are as given in Table II. From these roots we have calculated the $a/R$- dependent results for energy production as shown in Table III.

It is seen that with increasing values of the string width, the amount of produced energy increases. This is in accordance with expectations. The increase is however relatively moderate as long as the string stays "thin" relative to the radius of the waveguide: In the range $10^{-5} < a/R < 10^{-2}$ the energy increases only by a factor of about 7.

\section{Conclusive remarks}

Let us list some key points of the results achieved:

(1)  The starting point in the analysis was the Gott-Hiscock exterior metric (5), given for the annular region $ a \leq r \leq R$. All three types of the classical modes were worked out: The TM, TE, and TEM modes. The calculation was based upon perfect conductor boundary conditions at $r=a, R$.

(2)  Quantum mechanically, only the TM modes were considered, corresponding to the restriction (12). The continuity conditions on the field component $E_z$  at the instant $t=t_0$ were Eqs. (48) and (50). The energy production was found to be strongly peaked in the lowest $(m,s)=(0,1)$ mode; cf. Eq. (65). The much lower energy production in the next mode, the (1,1) mode, was given in Eq. (66). This expression shows the characteristic property of being proportional to $(G\mu/t)^2$; this is the same kind of behavior as found in the same mode in \cite{brevik95} and also for the scalar field in the azimuthally symmetric mode in \cite{parker87}.

(3)  The mathematical reason for the differences from the zero-width case lies in the need to include the Neumann function in the solutions for the mode functions; cf. Eqs. (43) and (15). The value of the energy produced was found to be greatly influenced by the roots of the boundary conditions, for $t<t_0$, and for $t>t_0$.

(4)  We assumed a GUT-scale string, and the input parameters (string mass, radius, waveguide dimensions), were chosen accordingly. The formalism of course permits scaling of these parameters. Inserting various values of the ratio $a/R$ we found that the value of the produced energy did not vary very much (less than about one order of magnitude) when the variations took place in the range allowed by the approximations used.

\renewcommand{\theequation}{\mbox{\Alph{section}\arabic{equation}}}

\appendix

\setcounter{equation}{0}

\section{On the evaluation of $I_{ss'}$}

In the expression (58) for $I_{ss'}$ we approximate $ q_{\beta ms}$ by $p_{ms}$, where the latter quantity is a root of $J_{|m|}(p_{ms}R)=0$. This corresponds to the second term in the condition (12) for the TM mode being negligible. Moreover, setting $\beta m \simeq m$ the integral in Eq. (58) becomes
\begin{equation}
\int_a^RJ_{|m|}(p_{ms}r)\mathcal{C}_{|m|}(p_{ms'}r)rdr=
\frac{1}{2}\left[r^2J_{|m|+1}(p_{ms}r)\mathcal{C}_{|m|+1}(p_{ms}r)\right]_a^R\delta_{ss'},
\end{equation}
where the orthogonality in $s$ stems from Eq. 11.3.29 in \cite{abramowitz72}. Remembering that $a/R \ll 1$ and making use of $ N_{|m|+1}(p_{ms}R)/N_{|m|}(p_{ms}a) \simeq 0$ as well as
\begin{equation}
\frac{N_{|m|+1}(z)}{N_{|m|(z)}}\simeq (|m|+1)\frac{2}{z}
\end{equation}
\label{A2}
(cf. Eq. 9.1.9 in \cite{abramowitz72}), we see that the contribution from the lower limit $r=a$ in Eq. (A1) can be omitted.

Moreover, the second factor in the denominator in Eq. (58) can be written
\begin{equation}
\left[r^2\{J_{|m|+1}(p_{ms}r)-\frac{J_{|m|}(p_{ms}a)}{Y_{|m|}(p_{ms}a)}Y_{|m|+1}(p_{ms}r)\}^2\right]_a^R.
\end{equation}
so that the expression (58) is approximately equal to 
\begin{equation}
I_{ss'}\simeq \frac{1}{2}\delta_{ss'},
\end{equation}
\label{A4}
as given in Eq. (59). As a corollary, we carried out numerical checks on the expression for $I_{ss'}$ using Maple V, supporting the result above.

\begin{table}[h] \label{table1}
\begin{center}
\begin{tabular}{lll}
\hline \hline
(m,s) & $x_{ms}$ & $x_{\beta ms}$ \\ \hline
(0,1) & 2.40483 & 2.54821 \\ 
(1,1) & 3.831711362 & 3.831711363 \\ 
(1,2) & 7.015592430 & 7.015592434 \\
(2,1) & 5.135632452 & 5.135632452 \\ 
\hline \hline
\end{tabular} \caption{Roots for various modes ($a/R = 10^{-5}$)}
\end{center}
\end{table}

\begin{table}[h] 
\begin{center}
\begin{tabular}{lll}
\hline \hline
a/R & $x_{\beta 01}$ & $x_{\beta 11}$ 	\\ \hline 
$10^{-1}$ & 3.313938715 & 3.940945907 	\\ 
$10^{-2}$ & 2.800921755 & 3.832889531 	\\ 
$10^{-3}$ & 2.654814168 & 3.831723171  	\\ 
$10^{-4}$ & 2.587120412 & 3.831711480 	\\ 
$10^{-5}$ & 2.548210049 & 3.831711363 	\\ 
\hline \hline 
\end{tabular} \caption{Roots for various values of a/R}
\end{center}
\end{table}

\begin{table}[h] \label{table2}
\begin{center}
\begin{tabular}{lll}
\hline \hline
a/R & $W_{01}\times 4LR$ & $W_{11}\times 4LR$ \\ \hline
$10^{-1}$ & 2,493918307$\times10^{-1}$ &	3.026897060$\times10^{-2}$ \\ 
$10^{-2}$ & 5.601176720$\times10^{-2}$ & 3.527890658$\times10^{-7}$ \\ 
$10^{-3}$ & 2.353818082$\times10^{-2}$ & 3.229962982$\times10^{-12}$ \\ 
$10^{-4}$ & 1.284358731$\times10^{-2}$ & 6.036824072$\times10^{-11}$ \\ 
$10^{-5}$ & 8.066977810$\times10^{-3}$ & 6.130061838$\times10^{-11}$ \\ 
\hline \hline
\end{tabular} \caption{Energy production in the modes (0,1) and (1,1)}
\end{center}
\end{table}

\end{document}